\documentclass[aps,apl,showpacs,amsfonts,amssymb,twocolumn]{revtex4}

\usepackage[dvips]{graphicx}

\begin{document}

\title {Photovoltaic Probe of Cavity Polaritons in a Quantum Cascade Structure}

\author{Luca Sapienza}
\email{luca.sapienza@univ-paris-diderot.fr}
\affiliation{Laboratoire Mat\'eriaux et Ph\'enom\`enes Quantiques,
Universit\'e Paris Diderot, Paris VII, 75205 Paris Cedex 13,
France}
\author{Raffaele Colombelli}
\affiliation{Institut d'Electronique Fondamentale, Universit\'e Paris Sud, CNRS, 91405 Orsay, France}
\author{Angela Vasanelli}
\affiliation{Laboratoire Mat\'eriaux et Ph\'enom\`enes Quantiques, Universit\'e Paris Diderot, Paris VII, 75205
Paris Cedex 13, France}
\author{Cristiano Ciuti}
\affiliation{Laboratoire Mat\'eriaux et Ph\'enom\`enes Quantiques, Universit\'e Paris Diderot, Paris VII, 75205
Paris Cedex 13, France}
\author{Christophe Manquest}
\affiliation{Laboratoire Mat\'eriaux et Ph\'enom\`enes Quantiques, Universit\'e Paris Diderot, Paris VII, 75205
Paris Cedex 13, France}
\author{Ulf Gennser}
\affiliation{Laboratoire de Photonique et Nanostructures, LPN-CNRS, Route de Nozay, 91460 Marcoussis, France}
\author{Carlo Sirtori}
\email{carlo.sirtori@univ-paris-diderot.fr}
\affiliation{Laboratoire Mat\'eriaux et Ph\'enom\`enes Quantiques,
Universit\'e Paris Diderot, Paris VII, 75205 Paris Cedex 13,
France}

\begin{abstract}

The strong coupling between an intersubband excitation in a
quantum cascade structure and a photonic mode of a planar
microcavity has been detected by angle-resolved photovoltaic
measurements. A typical anticrossing behavior, with a vacuum-field
Rabi splitting of 16~meV at 78K, has been measured, for an
intersubband transition at 163~meV. These results show that the
strong coupling regime between photons and intersubband
excitations can be engineered in a quantum cascade opto-electronic
device. They also demonstrate the possibility to perform
angle-resolved mid-infrared photodetection and to develop active
devices based on intersubband cavity polaritons.
\end{abstract}

\maketitle

Cavity polaritons are quasi-particles resulting from the strong
coupling between a confined electromagnetic field and a material
elementary excitation. They are the normal modes of the
light-matter Hamiltonian and show a typical energy anticrossing
behavior as a function of the energy detuning between the bare
photon mode and the material excitation.\cite{Weisbuch} The
minimum energy splitting, measured at resonance, is the so-called
vacuum-field Rabi splitting. In 2003, the first observation of
intersubband (ISB) polaritons in reflectivity measurements was
reported\cite{Dini}, as a result of the coupling between a
two-Dimensional Electron Gas (2DEG) excitation and a photon mode
in a planar microcavity based on total internal
reflection.\cite{A.Liu} In the same year, ISB polaritons were also
observed in bound to quasi-bound transitions in a quantum well
infrared photodetector.\cite{Liu} Because of the large oscillator
strength and of the relative low energy of the ISB transitions,
ISB polaritons are good candidates to explore a new regime of
light-matter coupling, where the Rabi frequency can be a
significant fraction of the transition frequency: the so-called
ultra-strong coupling regime.\cite{Cristiano} It is predicted that
quantum electrodynamics phenomena reminiscent of the dynamic
Casimir effect can be observed in this regime.\cite{Cristiano2,
Simone} Furthermore, recent works show a growing interest in
implementing ISB polaritons in devices \cite{Raffaele} and the
possibility of electrical control and ultrafast modulation of the
ISB strong coupling regime.\cite{Aji, Aji2}

In this letter, we report on the experimental observation of ISB cavity polaritons in angle-resolved
photovoltaic measurements, performed on a quantum cascade (QC) structure embedded in a planar microcavity
(Fig.1). The QC structure is based on a GaAs/Al$_{0.45}$Ga$_{0.55}$As heterostructure, grown by molecular-beam
epitaxy on an undoped GaAs (001) substrate. The planar microcavity, designed for the confinement of transverse
magnetic (TM) polarized radiation, is realized by sandwiching the QC structure between a low refractive index
Al$_{0.95}$Ga$_{0.05}$As layer and a top metallic mirror. In order to measure the photogenerated voltage,
circular mesas of 220~$\mu$m diameter are etched down to the bottom n-doped layer just below the active region.
Metallic contacts are then provided on the top of the mesa and on the n-doped layer (Fig.1a), so that electrons
can be extracted directly from the structure without having to cross the Al$_{0.95}$Ga$_{0.05}$As mirror. The
sample is soldered onto a copper holder and mounted on the cold finger of a cryostat, cooled at liquid Nitrogen
temperature. Angle-resolved measurements are performed by rotating the cold finger in order to probe cavity
modes with different energies. The sample facet is polished at $70\,^{\circ}$, in order to allow a variation of
the propagating angle of the incident beam over an angular range (of about $26\,^{\circ}$ internal angle) useful
to map the anticrossing curve.

\begin{figure}[h] \includegraphics[angle=0, width=0.35\textwidth]{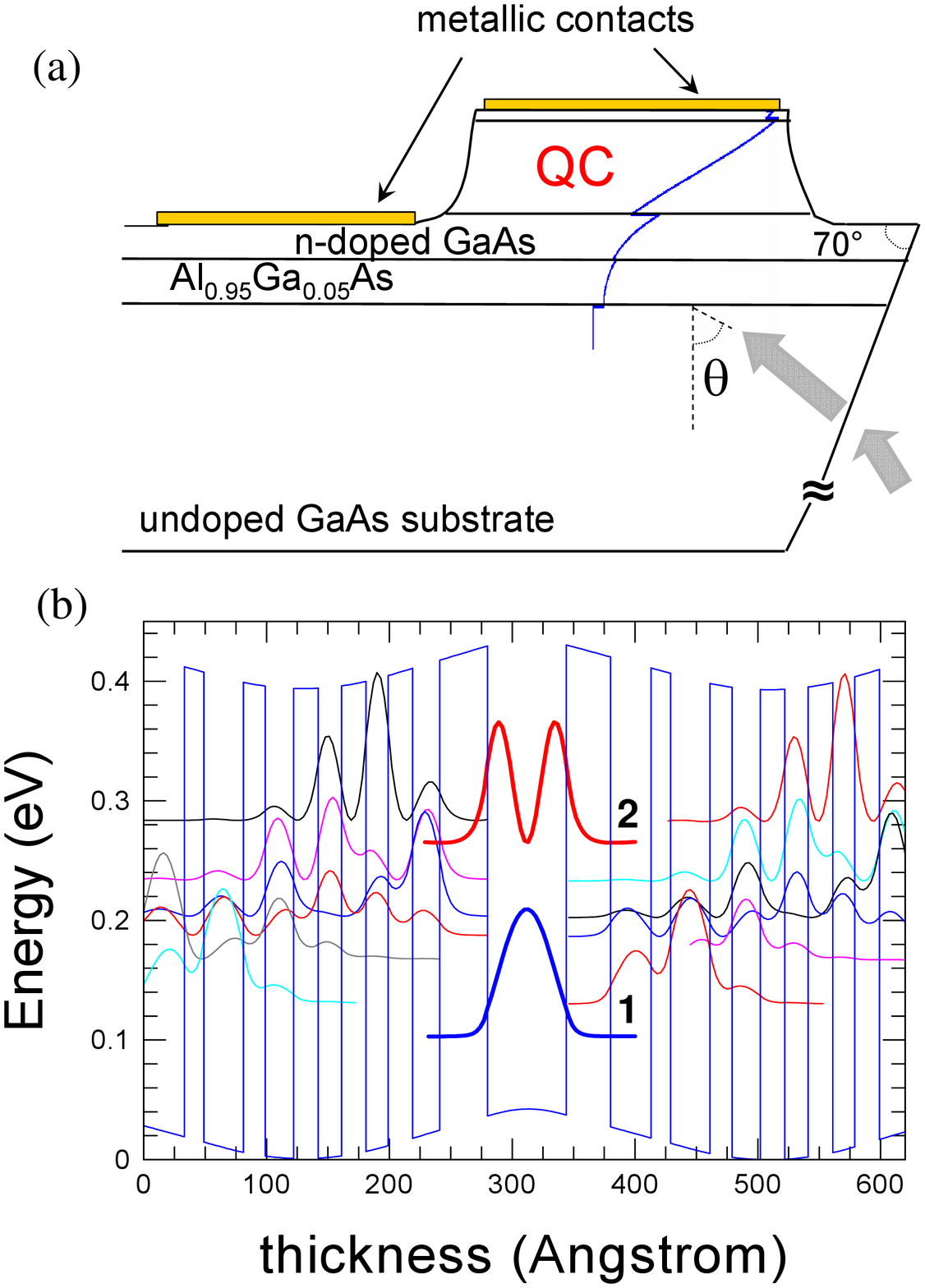}
\centering \caption{(a) Schematic view of the mesa etched sample.
The layers (Si-doping, thickness) from bottom to top are:
Al$_{0.95}$Ga$_{0.05}$As (undoped, 0.52~$\mu$m), GaAs
($3\times10^{18}$~cm$^{-3}$, 0.56~$\mu$m), QC structure, GaAs
($1\times10^{17}$~cm$^{-3}$, 86~nm), GaAs
($3\times10^{18}$~cm$^{-3}$, 17~nm), metallic contacts
[Ni(10nm)/Ge(60nm)/Au(120nm)/Ni(20nm)/Au(200nm)]. The arrows
represent the optical path of the incident beam. The intensity of
the photon mode along the growth axis is sketched (blue curve).
(b) Band diagram of the QC structure. The layer sequence of one
period of the structure, in nm, from left to right, starting from
the largest well is
$6.4/\textbf{3.6}/3.3/\textbf{1.6}/\underline{3.2/\textbf{1.8}/2.3/\textbf{2.0}/1.9/\textbf{2.0}}/1.8/\textbf{2.0}/2.2/\textbf{3.9}$.
Al$_{0.45}$Ga$_{0.55}$As layers are in bold, underlined layers are
$n$ doped with Si ($3\times10^{17}$~cm$^{-3}$). This sequence is
repeated 30 times.}
\end{figure}

The band diagram of the QC structure, obtained with
self-consistent Schr\"odinger-Poisson calculations, is presented
in Fig.1b. Photons confined within the microcavity can be absorbed
promoting electrons from level 1 to level 2 of the quantum well
($E_{21}=163$~meV), with the consequent creation of ISB
excitations in the 2DEG. The electronic band-structure engineering
of this QC structure, analogous to that of a quantum cascade
detector,\cite{QCD} is such that electrons preferentially scatter
towards one side (left side in Fig.1b) of the quantum well, giving
rise to a photovoltage.

\begin{figure}[h]\includegraphics[angle=0, width=0.35\textwidth]{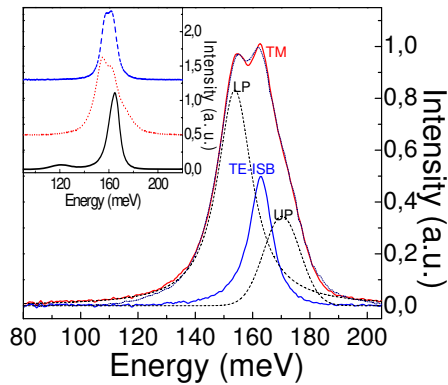}
\centering \caption{Normalized photovoltage spectra (spectral
resolution of 8 cm$^{-1}$, at 78K) for $\theta=75\,^{\circ}$. The
solid lines labelled with TM and TE-ISB represent the spectra
collected with TM- and TE-polarized light respectively. The two
peaks in dashed lines, labelled $LP$ and $UP$, are the Lorentzian
and Gaussian functions, respectively, used in the fitting
procedure explained in the text. The result of the three curve fit
is shown by the dotted line. Inset: photovoltage spectra with
different incident angles ($67.7\,^{\circ}$ solid line,
$76.3\,^{\circ}$ dotted line, $81.2\,^{\circ}$ dashed line).}
\end{figure}

In the experiment, the facet of the sample is illuminated with the
radiation of a Globar lamp, focused using a $f/1.5$ ZnSe
plano-convex lens. The spectra are collected using a Nicolet
Fourier-transform infrared spectrometer, in rapid scan mode. The
inset of Fig.2 shows three spectra collected with TM-polarized
incident light at three different angles, $67.7\,^{\circ}$,
$76.3\,^{\circ}$ and $81.2\,^{\circ}$. We observe a considerable
variation of the shape of the spectra as a function of the angle.
In Fig.2 the solid lines are two spectra obtained with TM- and
transverse electric (TE) polarized incident light with a
propagating angle of $75\,^{\circ}$ with respect to the growth
axis. The height of the TE curve has been adjusted following a
fitting procedure that we describe later in the text. Note that,
the TE spectrum is about 20 times weaker than the TM one. Even
though the polarization selection rule predicts that TM
polarization only can excite ISB transitions, we observe a
spectrum from the TE polarized light, possibly because of
scattering processes, occurring at the mesa sidewalls, which
randomize the momentum and the polarization of the incident
light.\cite{scattering} This peak has the same energy position and
shape in all the spectra collected at different angles and
corresponds to the bare ISB transition. Moreover, the TE spectrum
corresponds in shape and energy position to what can be measured
very far from the anticrossing region (e.g., at $30\,^{\circ}$)
and its energy position is in good agreement with the theoretical
energy difference $E_{21}$. In fact, the TE incident light cannot
be injected into the cavity mode, which is based on a surface
plasmon mode (Fig.1a). The spurious TE signal that we observe must
be produced by photons that have undergone a scattering event and
therefore have lost their original wavevector. Hence, we are in
the condition to measure the TM and the TE spectra for each angle
and to subtract one another to allow a better visualization and
analysis of the polaritonic contributions. This procedure has been
used to obtain the spectra in Fig.3a, collected for different
incident angles of the beam, at 78K. A clear anticrossing between
the cavity mode and the ISB excitation is visible, with a
vacuum-field Rabi splitting of 16meV.

\begin{figure}[h]\includegraphics[angle=0, width=0.35\textwidth]{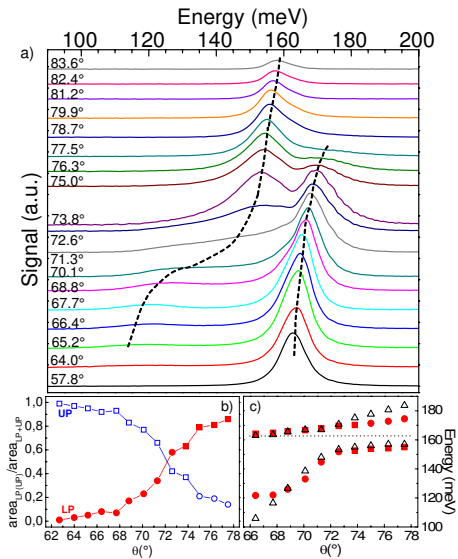}
\centering \caption{a) Photovoltage spectra (at 78~K) obtained
with TM-polarized incident radiation, after subtraction of the TE
peak, presented as a function of the propagating angle within the
cavity. The spectral resolution is 8~cm$^{-1}$ and the spectra are
offset each other for clarity. The dashed lines are guides for the
eye. b) Areas of the $UP$ (open symbols) and $LP$ (full symbols)
peaks, normalised to the total area, as a function of the internal
angle. Squares (circles) represent values obtained using a
Lorentzian (Gaussian) function in the fit. c) Energy position of
the photovoltage peaks as a function of the incident angle of the
radiation (squares for Lorentzian functions, circles for Gaussian
functions), compared with the results of the transfer matrix
calculations (open triangles). The dashed line shows the energy of
the bare ISB transition.}
\end{figure}

To explain our subtracting approach, we concentrate on the TM spectrum of Fig.2. We can distinguish the presence
of three peaks: the Lower Polariton ($LP$, dashed line), a middle peak ($TE-ISB$) and the Upper Polariton ($UP$,
dashed line). A three curve fitting procedure using a Lorentzian function (for the $LP$), a Gaussian function
(for the $UP$) and the TE-spectrum has been used to identify the height of the $TE-ISB$ peak. The use of two
different functions to obtain the best fit is due to the fact that, for the angle considered ($75\,^{\circ}$),
the $LP$ is still more ISB-like and therefore Lorentzian while the $UP$ is still more cavity-like, thus
Gaussian. The fitting curve is the dotted line and shows an excellent agreement with the experimental data. Note
that the Lorentzian and Gaussian shape swap at resonance in accordance with the \emph{light} and \emph{matter}
weight of the polaritonic wavefunctions (Fig.3b and c). The result of the fit indicates that the area of the
spurious $TE-ISB$ peak is $1/3$ of the total area at all angles. This suggests that the bare ISB signal is
proportional to the polaritonic contribution, as if it were a consequence of light scattered after polariton
absorption.

This observation motivates us to underline the main differences between our experiment and the absorption
experiments in which the bare ISB transition is absent.\cite{Dini, Aji} In absorption, the sample is unprocessed
and the wavevector of both incident and collected photons are highly selected, by using lenses with long focal
length. In our measurements, only the incident angle can be controlled. This implies that all the light
undergoing scattering processes within the cavity can contribute to the photovoltaic signal. The reproducibility
of our data has been verified on several devices and the presence of the middle peak has been observed also in
different processing geometries (ridges of different size). The presence of a third middle peak has already been
observed in photoluminescence measurements of exciton polaritons with a fluctuating environment.\cite{Hennessy}
Further experiments and theoretical investigations are in progress to clearly identify the feature in our
system.

In Fig.3b, the areas of the two polaritonic peaks, normalised to the total area, are plotted as a function of
the internal angle: the mixing of the ISB excitation and of the photon mode is evident. As expected, the upper
and lower polariton show the same area at zero detuning, when the photonic and the matter fractions are equal.
In Fig.3c, the position of the peaks as a function of the propagating angle (full symbols) is presented, as well
as the results of transfer matrix simulations (open triangles). To reproduce the experimental data, in the
theoretical calculations, which take into account the dispersion of the refractive index of the Au on the
surface (the Ni/Ge/Au alloy has been modeled as Au only), the electronic density has been set to
$1.8\times10^{11}$~cm$^{-2}$, a lower value than the nominal one. In the transfer matrix calculations, the ISB
excitation in the 2DEG has been taken into account by including in the dielectric constant of the quantum well
layers, an additional term in the form of an ensemble of classical polarized Lorentz oscillators.\cite{Dini,
Raffaele}

In conclusion, the results herein presented show that ISB transitions in QC structures can be used to achieve an
electrical read-out of states in the strong coupling regime. A signal is detected by the device only for photons
impinging with a well-defined angle, allowing an angle-resolved photodetection. This represents an electrical
probe of microcavity optical dynamics, with the potentiality to become a new important tool for the study of
cavity quantum electrodynamics in solid state systems.

\begin{acknowledgments}
We would like to acknowledge J.~Faist, Y.~Chassagneux, I.~Sagnes,
L. Largeau and O. Mauguin for help and useful discussions. We are
particularly grateful to S.~S.~Dhillon for his help and support in
the experiments. The device fabrication has been performed at the
nano-center "Centrale Technologique Minerve" at the \emph{Institut
d'Electronique Fondamentale}. We gratefully acknowledge support
from EU MRTN-CT-2004-51240 POISE and ANR-05-NANO-049-01 INTERPOL.
\end{acknowledgments}

\end{document}